\PassOptionsToPackage{table}{xcolor}

\documentclass[oneside,12pt,letterpaper]{article}

\usepackage{authblk} %

\usepackage{amsmath,amssymb,bm,bbold}
\usepackage{amsfonts,amsthm,mathrsfs,mathtools}
\usepackage{stmaryrd}
\SetSymbolFont{stmry}{bold}{U}{stmry}{m}{n}
\usepackage{upgreek}
\usepackage{stmaryrd}
\usepackage{econometrics}
\usepackage{flexisym}
\usepackage{breqn}

\usepackage{tikz}
\usepackage{xr-hyper}
\usepackage{pdfpages}
\usepackage{pgf,interval}

\usepackage{float}
\usepackage{booktabs, array, longtable, ragged2e, pdflscape}
\usepackage{subcaption}
\usepackage{tabularx}
\usepackage{multirow}
\newcolumntype{C}{>{\centering\arraybackslash}X}

\usepackage{hyperref}
\hypersetup{
  colorlinks = false,
  urlbordercolor = {blue},
  linkbordercolor = {red},
  citebordercolor = {green}
}
\usepackage[page,toc,titletoc,title]{appendix}
\usepackage{makeidx}
\usepackage[nottoc,numbib]{tocbibind}
\hypersetup{bookmarksopen=true}

\usepackage{cleveref}
\renewcommand{\cref}{\Cref}

\AtBeginEnvironment{appendices}{\crefalias{section}{appendix}}

\usepackage{algorithm}
\usepackage{algorithmic}
\usepackage[T1]{fontenc}
\usepackage[american]{babel}
\usepackage{csquotes}
\usepackage[
  style=authoryear,maxnames=4,natbib=true,uniquename=false,
  isbn=false,url=false,eprint=false,date=year,dashed=false
]{biblatex}
\bibliography{florida.bib}
\DefineBibliographyStrings{english}{bibliography = {References}}

\usepackage{xcolor}
\usepackage{cancel} %
\usepackage[left=2.5cm, right=2.5cm, top=3cm]{geometry}
\usepackage{indentfirst}
\usepackage{fancyhdr}
\makeatletter
\let\save@mathaccent\mathaccent
\newcommand*\if@single[3]{%
  \setbox0\hbox{${\mathaccent"0362{#1}}^H$}%
  \setbox2\hbox{${\mathaccent"0362{\kern0pt#1}}^H$}%
  \ifdim\ht0=\ht2 #3\else #2\fi
  }
\newcommand*\rel@kern[1]{\kern#1\dimexpr\macc@kerna}
\newcommand*\widebar[1]{\@ifnextchar^{{\wide@bar{#1}{0}}}{\wide@bar{#1}{1}}}
\newcommand*\wide@bar[2]{\if@single{#1}{\wide@bar@{#1}{#2}{1}}{\wide@bar@{#1}{#2}{2}}}
\newcommand*\wide@bar@[3]{%
  \begingroup
  \def\mathaccent##1##2{%
    \let\mathaccent\save@mathaccent
    \if#32 \let\macc@nucleus\first@char \fi
    \setbox\z@\hbox{$\macc@style{\macc@nucleus}_{}$}%
    \setbox\tw@\hbox{$\macc@style{\macc@nucleus}{}_{}$}%
    \dimen@\wd\tw@
    \advance\dimen@-\wd\z@
    \divide\dimen@ 3
    \@tempdima\wd\tw@
    \advance\@tempdima-\scriptspace
    \divide\@tempdima 10
    \advance\dimen@-\@tempdima
    \ifdim\dimen@>\z@ \dimen@0pt\fi
    \rel@kern{0.6}\kern-\dimen@
    \if#31
      \overline{\rel@kern{-0.6}\kern\dimen@\macc@nucleus\rel@kern{0.4}\kern\dimen@}%
      \advance\dimen@0.4\dimexpr\macc@kerna%
      \let\final@kern#2%
      \ifdim\dimen@<\z@ \let\final@kern1\fi
      \if\final@kern1 \kern-\dimen@\fi
    \else
      \overline{\rel@kern{-0.6}\kern\dimen@#1}%
    \fi
  }%
  \macc@depth\@ne%
  \let\math@bgroup\@empty \let\math@egroup\macc@set@skewchar%
  \mathsurround\z@ \frozen@everymath{\mathgroup\macc@group\relax}%
  \macc@set@skewchar\relax
  \let\mathaccentV\macc@nested@a%
  \if#31
    \macc@nested@a\relax111{#1}%
  \else
    \def\gobble@till@marker##1\endmarker{}%
    \futurelet\first@char\gobble@till@marker#1\endmarker%
    \ifcat\noexpand\first@char A\else%
      \def\first@char{}%
    \fi
    \macc@nested@a\relax111{\first@char}%
  \fi
  \endgroup
}
\makeatother

\usepackage{enumitem}

\usepackage{pifont}%
\newcommand{\interior}[1]{%
  {\kern0pt#1}^{\mathrm{o}}%
}

\numberwithin{theorem}{section}

\title{Migration of Voters in Florida, 2017-2022}
\author{Fangzhou Xie}
\affil{Department of Economics, Rutgers University\\
Center for Health Services Research, Rutgers University}
\date{\today}

\begin{document}
\maketitle

\begin{abstract}
  Florida has experienced significant population increase in recent years, driven in part by domestic migration from other states. This study analyzes the migration patterns of voters in Florida between 2017 and 2022 using voter registration data. By examining demographic characteristics such as race/ethnicity, gender, age, and party affiliation, I identify trends in voter migration and their implications for Florida's political landscape. The findings reveal that minorities, younger individuals, Republicans, and those possibly with non-conforming gender are more likely to migrate into Florida. These insights contribute to understanding the dynamics of Florida's migration patterns and the effect of migration on recent elections.
\end{abstract}

\tableofcontents
\newpage

\section{Introduction}

In the 2024 US presidential election, Donald Trump won Florida by 13.1\%,
solidifying the Sunshine State as a Republican stronghold.
Not too long ago in 2000, Florida was still a key battleground state,
with George W. Bush winning a tight margin after the recount dispute.
The demographic changes in Florida over the past two decades
have played a significant role in shaping its political landscape.
Understanding these changes is crucial for political strategists,
policymakers, and researchers interested in electoral dynamics.

Leveraging the Florida Voter Registration files of 2017,
\citet{xie2021,xie2022} developed machine learning models
to infer race/ethnicity only from people's names.
Those voter registration files were later updated again in 2022
\citep{sood2022},
providing a unique opportunity to analyze the migration patterns
of voters in Florida over a five-year period.

In this paper, I analyze the migration patterns of voters in Florida
between 2017 and 2022 using Florida voter registration data \citep{sood2022}.
Specifically, I examine the voters who moved into and out of Florida
during this period and analyze the net migration patterns
stratified by demographic variables.
Moreover, as the voter registration files contain individual-level
address information, I further analyze the geographic patterns
of voter migration at the county level.

This paper specifically contributes to three strands of literature:
first, on migration and political change, especially given the
data on voter registration;
second, on the effects of demographic changes on voting patterns;
third, on migration patterns
among elderly populations, as Florida is a well-known retirement destination.

First, \citet{jurjevich2012} studied the impact of migration
on political change in both origin and destination states.
Moreover, the party affiliation of Florida migrants
has also changed over time with the national realignment of party coalitions.
In the 1970s and 1980s, White migrants to the South tended to be Republican
\citep{hillygus2017}, and they are more likely to be younger, better educated,
and have higher incomes \cite{brown2018}.
However, in the 2000s, migrants to Florida were increasingly
Democratic-leaning \citep{hillygus2017,mckee2016},
even leading to Obama winning Florida in 2008 and 2012.
In recent years, however, newcomers to Florida have been more aligned
with the GOP, especially after Ron DeSantis became governor in 2018
\citep{mckee2025}.
In this paper, I offer some insights into the migration patterns of voters in Florida,
and their implications for the political landscape in Florida,
especially as the Sunshine State has come to be regarded as a reliable Republican stronghold.

Second, there are abundant studies on the effects of demographic changes on voting patterns,
for example, White vs non-White voters showing different preferences for the racially
relevant initiatives \citep{branton2004},
racial and ethnic diversity is associated with higher levels of support for
Hillary Clinton in 2016, but not for anti-immigration candidates \citep{hill2019}.
Apart from the voting differences between different racial and ethnic groups,
there are also gender-differential effects on voting outcomes.
Although traditionally younger voters are more liberal than older ones \citep{geys2022},
there has been a recent trend of younger male voters voting for the right-wing parties.
This has been documented in recent 2024 European elections \citep{milosav2026},
and also predicted by a model in the US
explaining the 2016 election results for the young white male voters \citep{ghitza2023}.

Third, it is well known that Florida is a popular retirement destination
\citep{walters2002},
especially among US-born retirees
\citep[``classic retirement magnet'']{frey2000}.
One reason could be the school tax and expenditure structure:
the elderly population tends to prefer areas with low public school expenditure
\citep{assadian1995},
which is in alignment with the Tiebout hypothesis \citep{tiebout1956},
a seminal theory in local public economics that explains how individuals choose
their residential location based on the provision of
public goods and services\footnote{
  ``Consider for a moment the case of the city resident about to move to the suburbs.
  What variables will influence his choice of a municipality?...
  The consumer-voter may be viewed as picking that community which best
  satisfies his preference pattern for public goods.'' --- \citet{tiebout1956}
}.
In this paper,
however, I find that retirees (aged 65+) are moving out of Florida,
which is a surprising trend that contradicts the common perception of Florida as a retirement haven.

Further,
leveraging the geographic information in the voter registration files
at the individual level, I can also stratify by
demographic variables and analyze migration patterns at the county level.
These migration patterns can be visualized using Florida
county maps, providing intuitive insights into the geographic distribution.

The rest of the paper is organized as follows:
\cref{sec:methods} describes the data used in the analysis,
the Florida Voter Registration files from 2017 and 2022,
and the summary statistics of the data.
\cref{sec:results} presents the main results of the migration analysis,
including the overall migration patterns, demographic stratifications,
and county-level geographic patterns.
\cref{sec:conclusion} concludes.

\section{Materials and Methods}\label{sec:methods}

Several studies of race and ethnicity prediction from names
\citep{xie2021,xie2022,chintalapati2023}
leverage the Florida Voter Registration files of 2017 \citep{sood2022},
as these files provide a rather unique dataset where both names and self-reported
race/ethnicity are available for more than 20 million individuals in Florida.
This gives researchers the opportunity to train and validate name-based
race/ethnicity imputation models on a large and diverse dataset.

Moreover, the dataset was updated again to contain the voter registration information
in 2022 \citep{sood2022}.
This update allows researchers to study the migration patterns of Florida voters
across this five-year period.
This also coincides with the Covid pandemic period,
where the sunshine state saw a large influx of new residents from other states
\citep{floridachamberofcommerce2024}.

More importantly,
the voter registration files contain the detailed geographical location
at the individual level, which would allow us to track the migration patterns
at the county level.
For example, in the following subsections
we can visualize the changes of voters from 2017 to 2022,
and the net change by comparing those two,
stratified by demographic variables, i.e.\ race/ethnicity, gender,
age group, and party affiliation.
Given the nature of the voter registration files being administrative data,
they provide an accurate and comprehensive record of the voting population,
as opposed to survey-based Census data.
Additionally, the voter registration files give us the unique identifiers
for each voter, which allows us to track the migration of voters across the
two years, not only to identify the voters who moved in and out of Florida,
but also those who moved within Florida, which is not possible with the Census data.
This advantage of the voter registration files enables us to directly track
the voting population, which is of more interest to political studies.

To identify the migration patterns of voters in Florida between 2017 and 2022,
I first compare the two voter registration files and identify the voters who only
appear in one of the two years, and those who appear in both years.
If a voter only appears in the 2022 file, but not in the 2017 file,
we can infer that this voter is a new resident who moved into Florida during this period;
whereas if a voter only appears in the 2017 file, but not in the 2022 file,
we can infer that this voter is a former resident who moved out of Florida during this period.
In this study, I refer to the former group as ``move-in'' voters,
and the latter group as ``move-out'' voters.
We can then stratify the move-in and move-out voters by their demographic characteristics,
and calculate the net move-in as the difference between the number of move-in voters and move-out voters.
For each of the stratification, we can also calculate the proportion of each group in the respective samples.

Further, as we also have the residential address information for each voter in the files,
we can also track the voter migration patterns at the county level,
and visualize the geographic distribution of the move-in and move-out voters across Florida counties.
By stratifying the migration patterns by demographic variables,
we can also analyze the migration patterns for different demographic groups across the counties in Florida.
This is achieved by counting the number of move-in and move-out voters in each county,
and then calculating the net move-in for each county and demographic group.
To standardize the data and make the migration patterns comparable across different counties
and strata, I calculate the county-level ratio by taking the value of each county and demographic group,
divided by the total number of voters among all counties and that demographic group in 2017.

Despite the advantages of using the voter registration files,
there are potentially some caveats in using this data to analyze the migration patterns of voters.
The first important caveat is that the changes in voter registration files
could also be due to the native Florida residents who newly registered
to vote (e.g. turning 18 years old during this period),
and there's no way for us to tell apart those new registrants from
the actual new residents who moved into Florida from other states.
However, as can be seen in \cref{tab:summary_inoutnet},
those local newly registered voters can at most affect the age group 18-24,
which accounts for only a relatively small proportion of the entire voter change.
Moreover, specifically for the age group 65+ as discussed in \cref{subsec:geo-age},
its migration patterns are not affected by local new registrants turning 18.

Another potential issue with this approach is that we cannot account for
mortality during this period, which would result in some voters appearing
in the 2017 file but not in the 2022 one.
This issue is particularly relevant for the older age groups, particularly those above 65,
and especially given the Covid pandemic during this period.
Due to the limitation of this dataset, we cannot directly account for
the mortality effect, but I attempt to address this issue by analyzing
the mortality and population rates in Florida during this period (\cref{subsec:geo-age}),
and examine the intra-state migration patterns to argue that
the migration patterns discussed in this paper are not driven by the mortality effect,
but rather the actual migration of voters in and out of Florida,
as well as within the state.

\cref{tab:summary_2017} and \cref{tab:summary_2022} list
the summary statistics of Florida voters in 2017 and 2022 respectively,
grouped by race/ethnicity\footnote{
  The race variable in the Florida Voter Registration files contains the following categories:
  American Indian or Alaskan Native, Asian or Pacific Islander,
  Non-hispanic Black, Hispanic, Non-hispanic White, Other, Multi-racial, and Unknown.
  The nine categories are reduced to 5 in this study, following \citet{xie2022}.
}, gender\footnote{
  The gender variable is coded in three categories: female, male, and unknown.
}, age group\footnote{
  The age groups are coded as: below 18, 18-24, 25-34, 35-44, 45-64, 65+.
  Notice that voters below 18 are not eligible to vote,
  and therefore should not appear in the voter registration files;
  however, the age variable is calculated using the birth year provided in the files,
  and the current year (2017 or 2022), and is thus only an approximation of the voters' actual age.
}, and party affiliation\footnote{
  The parties listed in the files are:
  American's Party of Florida (AIP), Constitution Party of Florida (CPF),
  Florida Democratic Party (DEM), Ecology Party of Florida (ECO),
  Green Party of Florida (GRE), Independence Party of Florida (IDP),
  Independent Party of Florida (INT),
  Libertarian Party of Florida (LPF), No Party Affiliation (NPA),
  Party of Socialism and Liberation - Florida (PSL),
  Reform Party of Florida (REF), and Republican Party of Florida (REP).
  This list is reduced to 4 categories: DEM, REP, NPA, and OTH (Other).
}.
In terms of race and ethnicity,
non-Hispanic White voters constitute the largest proportion of Florida voters,
with more than 60\% for both 2017 and 2022,
and higher in 2017 than in 2022.
Hispanic population is the second largest group,
and also saw an increase in proportion from 16\% in 2017 to 18\% in 2022.
Black voters remain largely the same proportion at around 14\% in both years.
Asian voters saw a small increase from 1.8\% to 2.1\%.

\begin{table}[H]
  \caption{Summary of Voter Demographics* in 2017.\label{tab:summary_2017}}
  \begin{tabularx}{\linewidth}{CCCC}
  \toprule
  \textbf{Group}                        & \textbf{Demographics} & \textbf{Counts} & \textbf{Ratio} \\
  \midrule
  \multirow[m]{5}{*}{Race}              & Asian                 & 241,233         & 0.0184         \\
                                        & Black                 & 1,792,527       & 0.1364         \\
                                        & Hispanic              & 2,126,859       & 0.1619         \\
                                        & White                 & 8,342,400       & 0.6349         \\
                                        & Other                 & 636,941         & 0.0485         \\
  \midrule
  \multirow[m]{3}{*}{Gender}            & Female                & 6,913,078       & 0.5261         \\
                                        & Male                  & 5,885,537       & 0.4479         \\
                                        & Other                 & 341,345         & 0.0260         \\
  \midrule
  \multirow[m]{5}{*}{Age Group}         & 18-24                 & 1,127,303       & 0.0858         \\
                                        & 25-34                 & 2,049,932       & 0.1560         \\
                                        & 35-44                 & 1,834,015       & 0.1396         \\
                                        & 45-64                 & 4,439,570       & 0.3379         \\
                                        & 65+                   & 3,689,140       & 0.2808         \\
  \midrule
  \multirow[m]{4}{*}{Party Affiliation} & DEM                   & 5,001,890       & 0.3807         \\
                                        & REP                   & 4,559,152       & 0.3470         \\
                                        & NPA                   & 3,305,151       & 0.2515         \\
                                        & OTH                   & 273,767         & 0.0208         \\
  \bottomrule
\end{tabularx}

  \noindent{\footnotesize{* Summary statistics of Florida voters in 2017,
      by race, gender, age group, and party affiliation.}}
\end{table}

\begin{table}[H]
  \caption{Summary of Voter Demographics* in 2022.\label{tab:summary_2022}}
  \begin{tabularx}{\linewidth}{CCCC}
  \toprule
  \textbf{Group}                        & \textbf{Demographics} & \textbf{Counts} & \textbf{Ratio} \\
  \midrule
  \multirow[m]{5}{*}{Race}              & Asian                 & 303,745         & 0.0211         \\
                                        & Black                 & 1,946,369       & 0.1353         \\
                                        & Hispanic              & 2,595,001       & 0.1804         \\
                                        & White                 & 8,737,394       & 0.6074         \\
                                        & Other                 & 801,777         & 0.0557         \\
  \midrule
  \multirow[m]{3}{*}{Gender}            & Female                & 7,496,981       & 0.5212         \\
                                        & Male                  & 6,490,099       & 0.4512         \\
                                        & Other                 & 397,206         & 0.0276         \\
  \midrule
  \multirow[m]{5}{*}{Age Group}         & 18-24                 & 1,119,321       & 0.0778         \\
                                        & 25-34                 & 2,203,391       & 0.1532         \\
                                        & 35-44                 & 2,116,189       & 0.1471         \\
                                        & 45-64                 & 4,661,269       & 0.3241         \\
                                        & 65+                   & 4,284,116       & 0.2978         \\
  \midrule
  \multirow[m]{4}{*}{Party Affiliation} & DEM                   & 5,033,237       & 0.3499         \\
                                        & REP                   & 5,127,109       & 0.3564         \\
                                        & NPA                   & 3,968,656       & 0.2759         \\
                                        & OTH                   & 255,284         & 0.0177         \\
  \bottomrule
\end{tabularx}

  \noindent{\footnotesize{* Summary statistics of Florida voters in 2022,
      by race, gender, age group, and party affiliation.}}
\end{table}

There appears to be gender gap in the Florida voter registration files.
Around 52\% of the registered voters in both 2017 and 2022 are female,
while only 45\% are male.
There are also around 3\% of the voters with unknown gender\footnote{
  The Florida Voter registration file documentation does not explain
  why some voters have unknown gender.
  According to the Florida voter registration form accessed online,
  there is a checkbox for the registrant to indicate their gender (M/F),
  and we can only assume that those with unknown gender
  either did not fill in the form correctly,
  or they chose not to fill in this box to identify themselves as a
  binary gender.
  We will further discuss this aspect in \cref{subsec:geo-gender}.
}.

In terms of the age distribution, the composition is relatively stable.
The largest age group is 45-64, with around 32-33\% in both years,
with the second largest group being 65+ around 28-29\%.
The youngest group 18-24 only accounts for around 7-8\% of the voters,
and the 25-34 group accounts for around 15\%.
The 35-44 group is also stable at 14\%.

Finally, in terms of party affiliation,
the Democratic Party (DEM) has a slight edge over the Republican Party (REP)
in 2017, but not anymore in 2022 with GOP voters being the largest group.
There also appear to be more voters registering as No Party Affiliation (NPA)
and for other parties (OTH) in 2022 compared to 2017.

Now, if we look at the voter files for both 2017 and 2022,
and then identify and remove everyone who appears in both years,
we can obtain two samples:
(1) those who appear only in 2022, but not in 2017,
which we call the ``move-in'' voters;
(2) those who appear only in 2017, but not in 2022,
which we call the ``move-out'' voters.
Further, for each demographic stratification,
we could count the number of each group
and calculate the ``net move-in'' as the difference
between ``move-in'' and ``move-out'' voters.
The results are shown in \cref{tab:summary_inoutnet}.
This will give us a better picture in terms of the changes
in the Florida voter registration files between 2017 and 2022.

From \cref{tab:summary_inoutnet}, we can see that Florida
saw a large influx of new voters during this 5-year period,
across almost all demographic groups.
Racially speaking,
the largest net increase came from Hispanic voters,
who constitute 37\% of the net increase.
The proportion of Asian net move-in voters is 5.5\%,
and higher than their share in the overall population
(2.1\% in 2022).
The Black population saw a net increase of 11.1\%,
which is slightly less than their share of 13.5\% in 2022,
while the White population only accounted for 27.4\% of the net increase,
which is much lower than their share of 60.7\% in 2022.

\begin{table}[H]
  \caption{Change of Voter Demographics* 2017-2022.\label{tab:summary_inoutnet}}
  \begin{tabularx}{\linewidth}{CCCC|CC|CC}
  \toprule
  \textbf{Group}                        & \textbf{Demographics} & \textbf{Counts In} & \textbf{Ratio In} & \textbf{Counts Out} & \textbf{Ratio Out} & \textbf{Counts Net} & \textbf{Ratio Net} \\
  \midrule
  \multirow[m]{5}{*}{Race}              & Asian                 & 105,391            & 0.027             & 38,079              & 0.014              & 67,312              & 0.054              \\
                                        & Black                 & 468,911            & 0.121             & 329,528             & 0.125              & 139,383             & 0.112              \\
                                        & Hispanic              & 781,266            & 0.202             & 323,674             & 0.123              & 457,592             & 0.369              \\
                                        & White                 & 2,163,054          & 0.559             & 1,813,608           & 0.690              & 349,446             & 0.282              \\
                                        & Other                 & 349,349            & 0.090             & 122,392             & 0.047              & 226,957             & 0.183              \\
  \midrule
  \multirow[m]{3}{*}{Gender}            & Female                & 1,840,381          & 0.476             & 1,305,797           & 0.497              & 534,584             & 0.431              \\
                                        & Male                  & 1,820,657          & 0.471             & 1,260,031           & 0.480              & 560,626             & 0.452              \\
                                        & Other                 & 206,933            & 0.053             & 61,453              & 0.023              & 145,480             & 0.117              \\
  \midrule
  \multirow[m]{5}{*}{Age Group}         & 18-24                 & 927,856            & 0.240             & 186,598             & 0.071              & 741,258             & 0.376              \\
                                        & 25-34                 & 636,032            & 0.164             & 447,854             & 0.170              & 188,178             & 0.095              \\
                                        & 35-44                 & 561,724            & 0.145             & 306,557             & 0.117              & 255,167             & 0.129              \\
                                        & 45-64                 & 1,038,940          & 0.269             & 616,354             & 0.235              & 422,586             & 0.214              \\
                                        & 65+                   & 703,419            & 0.182             & 1,069,918           & 0.407              & -366,499            & 0.186              \\
  \midrule
  \multirow[m]{4}{*}{Party Affiliation} & DEM                   & 1,125,912          & 0.291             & 1,013,674           & 0.386              & 112,238             & 0.090              \\
                                        & REP                   & 1,265,489          & 0.327             & 884,341             & 0.337              & 381,148             & 0.307              \\
                                        & NPA                   & 1,356,428          & 0.351             & 664,957             & 0.253              & 691,471             & 0.557              \\
                                        & OTH                   & 120,142            & 0.031             & 64,309              & 0.024              & 55,833              & 0.045              \\
  \bottomrule
\end{tabularx}

  \noindent{\footnotesize{* The sample of ``move-in'' voters refer to those who show up in the 2022
      file, but not in 2017 file; the sample of ``move-out'' voters refer to those who show up in the 2017 file, but not in 2022 file. Because all voters are uniquely identified by their voter ID numbers, we can track the migration of voters across the two years.
      Further, the ``net-move'' is calculated as the difference between the number of move-in voters and move-out voters within each demographic group.}}
\end{table}

Although we have noticed the gender gap in the overall voter registration
in both 2017 and 2022,
the move-in and move-out voters appear to have a more balanced
gender composition.
The move-in female and male voters are 47.6\% and 47.1\% respectively,
while the move-out female and male voters are 49.7\% and 48.0\% respectively.
The net move-in voters are slightly more males (45.1\%) than females (42.9\%),
but we also see an even larger proportion of the unknown gender (11.9\%).

The age distribution of the relocated voters shows quite an
interesting pattern:
the majority of the net move-in voters are from the
18-24 (36.7\%) and 45-54 (21.7\%) age groups,
while 25-34 and 35-44 age groups only account for 9.7\% and 13.1\%.
Most surprisingly, the 65+ age group
saw a net move-out of voters during this period,
accounting for 18.8\% of the net change.
Florida has been known for being a popular retirement destination,
for example in 2000 alone almost 400,000 people above age 60 moved to Florida,
consisting of 19\% of all 60+ migrants in the US that year,
and the top 1 destination with the second being California,
which only attracted 6.5\% \citep{longino2003}.

Finally, in terms of party affiliation,
the majority of net move-in voters did not identify their
party affiliation, accounting for 56.3\% of the net increase.
Net move-in Republican voters account for 30.6\% of the net increase,
whereas Democrats constitute only 8.5\% of the net increase.
This observation is consistent with ongoing changes in the political
landscape in the United States,
with Florida no longer being seen as a swing state,
but rather a reliable Republican stronghold in recent years,
especially in the 2024 presidential election
where Donald Trump won Florida by a margin of 13 percentage points
\citep{270towin2024,anderson2024}.

\section{Results}\label{sec:results}

\begin{figure}[H]
  \centering
  \includegraphics[width=\linewidth]{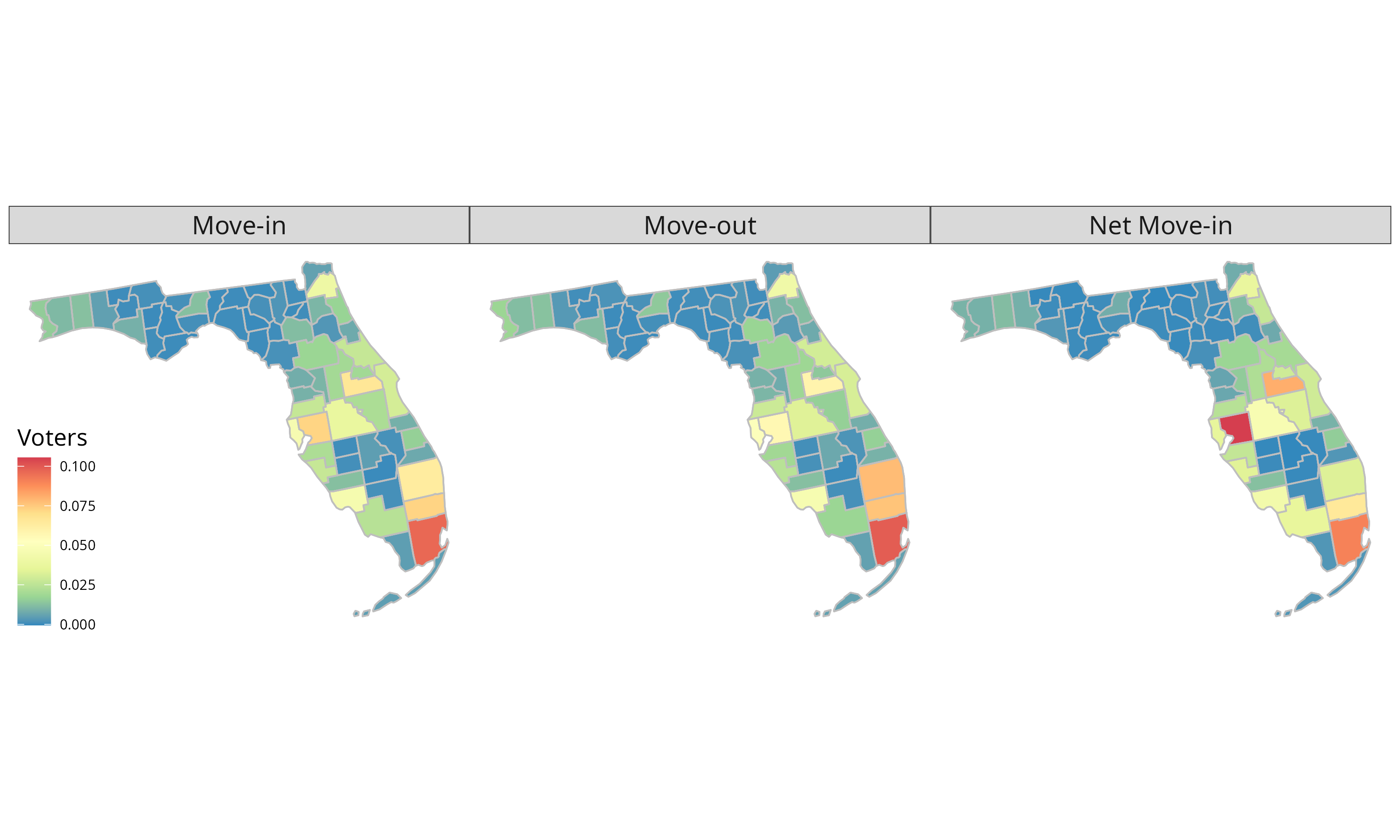}
  \caption{Ratio of move-in, move-out, and net move-in voters by county.\label{fig:map-overall}}
\end{figure}

We can further leverage the geographical information of the voters
to track migration patterns across different counties in Florida.
\cref{fig:map-overall} visualizes the overall ratio of move-in, move-out,
and net move-in voters by county.
From the maps, we can see that the changes in the populations
happened mostly in the urban areas, such as Miami-Dade county,
Broward county, Palm Beach county in South Florida,
Tampa Bay area (Hillsborough county, Pinellas county),
Orange county (Orlando area) in Central Florida,
and Duval county (Jacksonville area) in North Florida.
This also confirms the reports from the 2020 Census
that ``almost 97\% of Florida's population lived in metropolitan areas
in 2020'' \citep{universityofflorida2021}.
Moreover, Hillsborough, Miami-Dade, and Orange counties
saw the largest net increase of voters during the 5-year period,
mostly due to the cities Tampa, Miami, and Orlando.

\subsection{Geographic Distribution of Voter Migration by Race/Ethnicity}\label{subsec:geo-race}

Given that the Florida Voter registration files contain detailed geographic
information at the individual level, we could use the zip codes of their residential addresses
to track the migration patterns of voters across different counties in Florida.
Specifically, for each county, we calculate the number of move-in voters
(voters who appear only in the 2022 file, but not in the 2017 file),
move-out voters
(voters who appear only in the 2017 file, but not in the 2022 file),
and net-move voters
(calculated as the difference between the number of move-in voters and move-out voters
within each county).
We can also stratify the counts by race within each county,
and calculate county-wise proportions of move-in, move-out, and net-move voters
by race.
For example, among all Asian move-in voters in Florida,
what proportion of them move into Miami-Dade county, Broward county, Palm Beach county, etc.?
Then we could visualize these proportions on maps of Florida counties.
The same procedure is applied to move-out voters and net-move voters,
as well as in all subsequent subsections.

\begin{figure}[H]
  \centering
  \includegraphics[width=\linewidth]{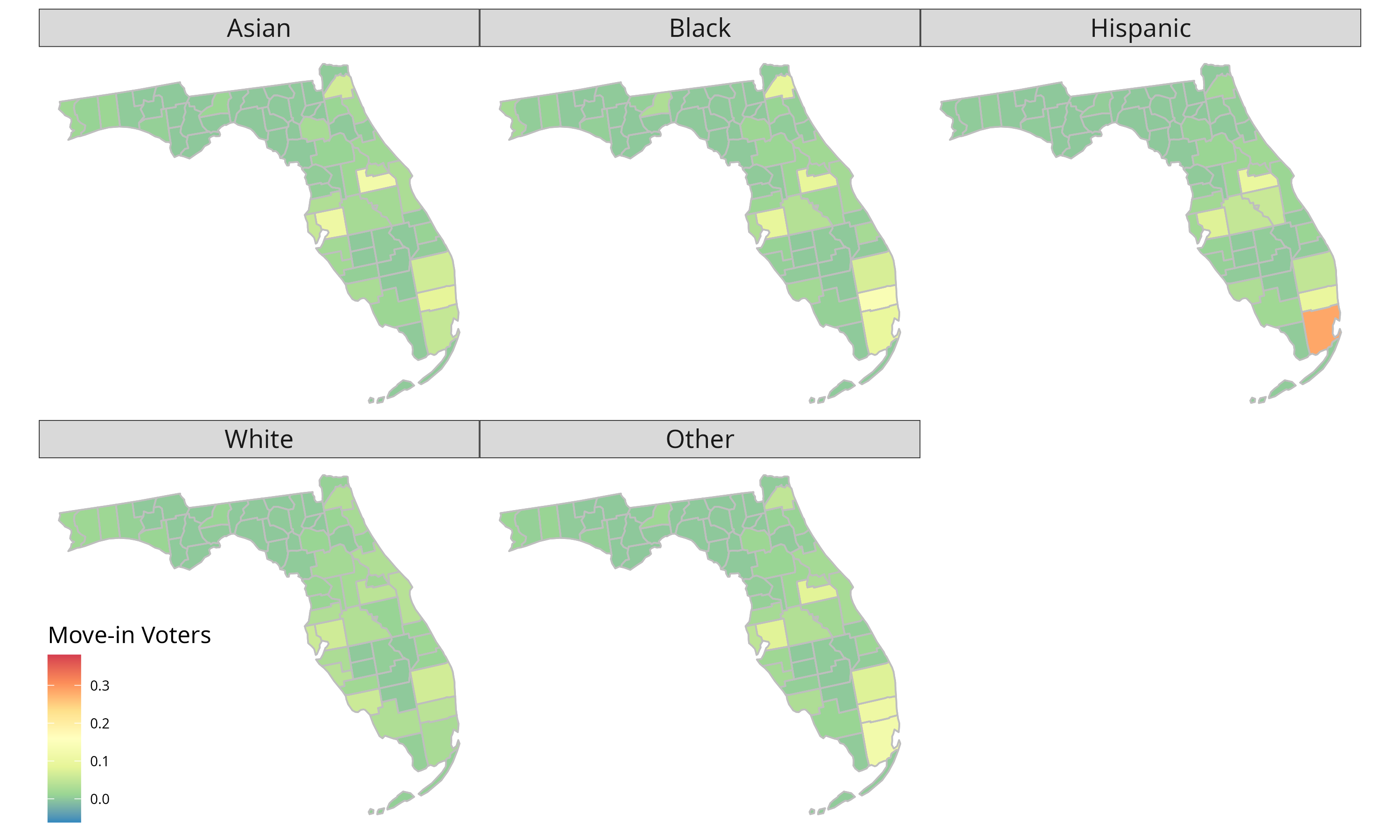}
  \caption{Ratio of move-in voters by race/ethnicity.\label{fig:map-race-in}}
\end{figure}

\begin{figure}[H]
  \centering
  \includegraphics[width=\linewidth]{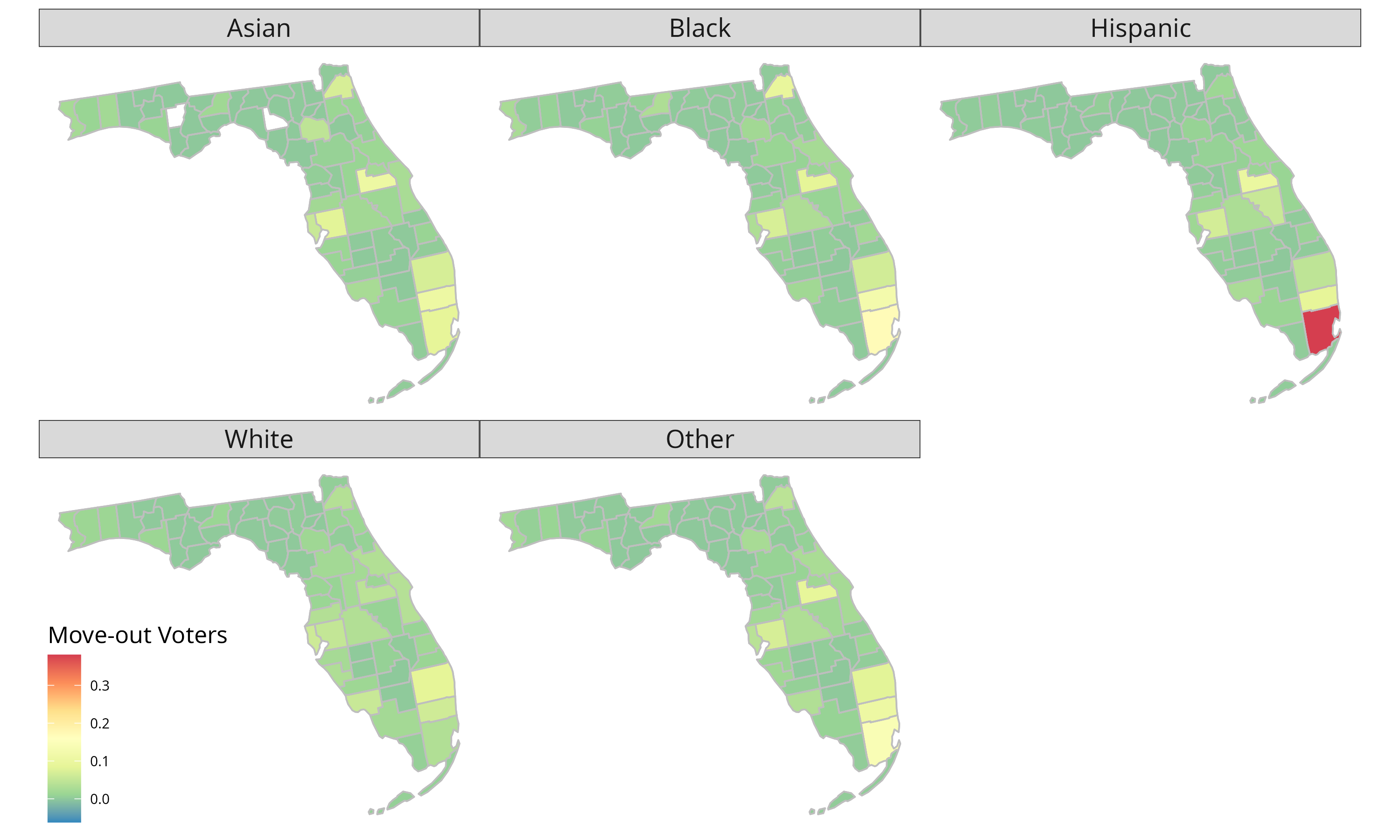}
  \caption{Ratio of move-out voters by race/ethnicity.\label{fig:map-race-out}}
\end{figure}

From \cref{fig:map-race-in},
we can tell that the move-in voters are more concentrated
for the minorities (Asian, Black, Hispanic, Other)
than the (non-Hispanic) White population.
For example, the minority populations are mostly moving
into the urban areas, such as Miami-Dade, Broward, Palm Beach,
Tampa Bay area (Hillsborough), Orange, and Duval county.
Hispanic populations are especially concentrated in
moving to the South Florida area (Miami-Dade),
and less so the Duval county,
compared to Asian and Black populations.
On the other hand, the (non-Hispanic) White population
are more spread out across all counties, with only mildly higher
proportions moving into the urban areas.
If we also look at \cref{fig:map-race-out},
we can see a similar pattern for the move-out voters.
Most notably, the majority of the moving-out Hispanic population
are from the Miami-Dade county.

\begin{figure}[H]
  \centering
  \includegraphics[width=\linewidth]{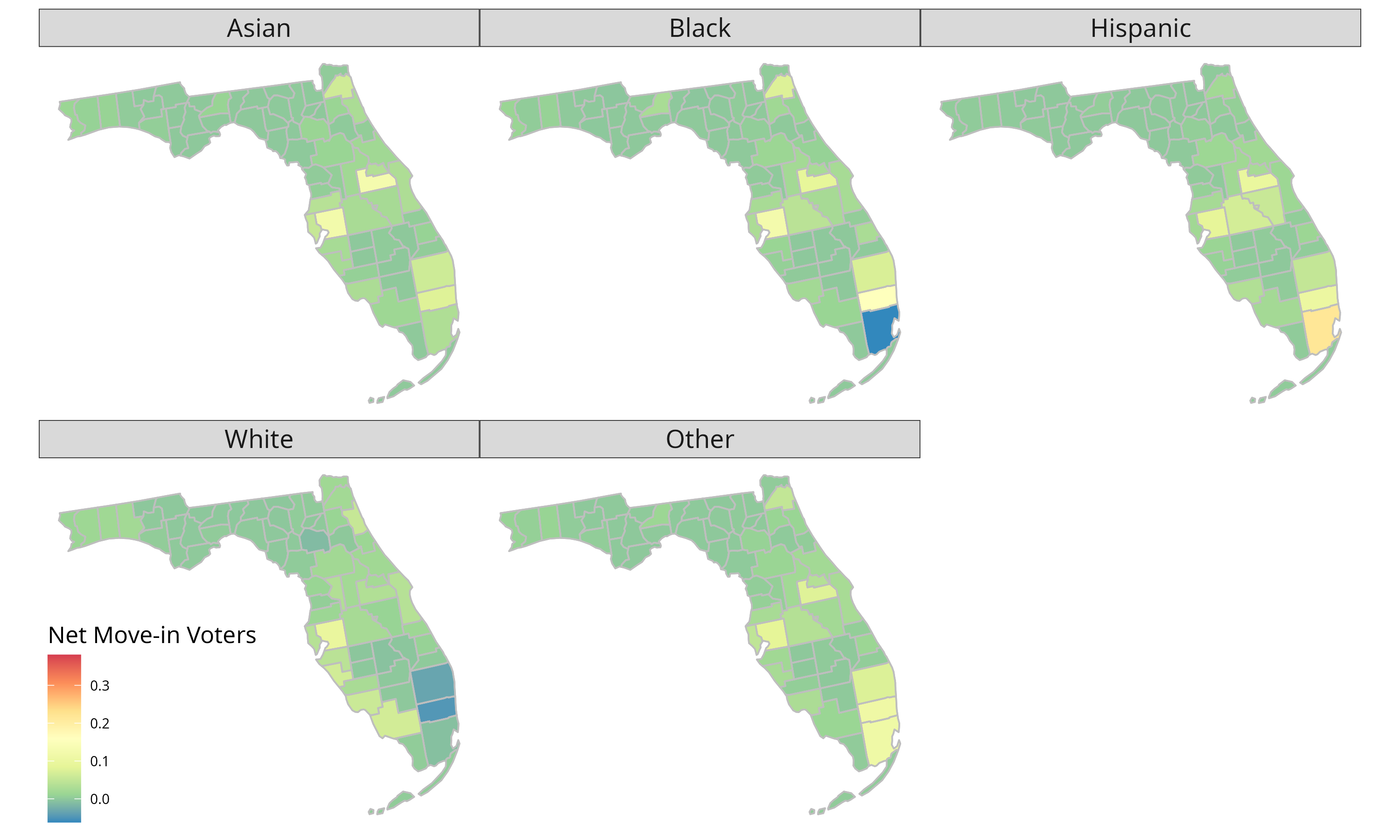}
  \caption{Ratio of net move-in voters by race/ethnicity.\label{fig:map-race-net}}
\end{figure}

Now, if we calculate the per-county net move-in voters stratified by race
in \cref{fig:map-race-net},
we can clearly see that even though Florida attracted a large number of
new voters into the urban areas (Miami region, Tampa Bay area, Duval county),
we have a net move-out of Black population in Miami-Dade county,
and a net move-out of White population in Broward and Palm Beach counties.
The rural counties did not appear to attract many new voters during this period,
and even saw a net move-out of voters in some counties for White population.

Geographically speaking, there is clearly a pattern of different
moving behaviors across different racial and ethnic groups,
especially between White versus non-White populations.

\subsection{Geographic Distribution of Voter Migration by Gender}\label{subsec:geo-gender}

\begin{figure}[H]
  \centering
  \includegraphics[width=\linewidth]{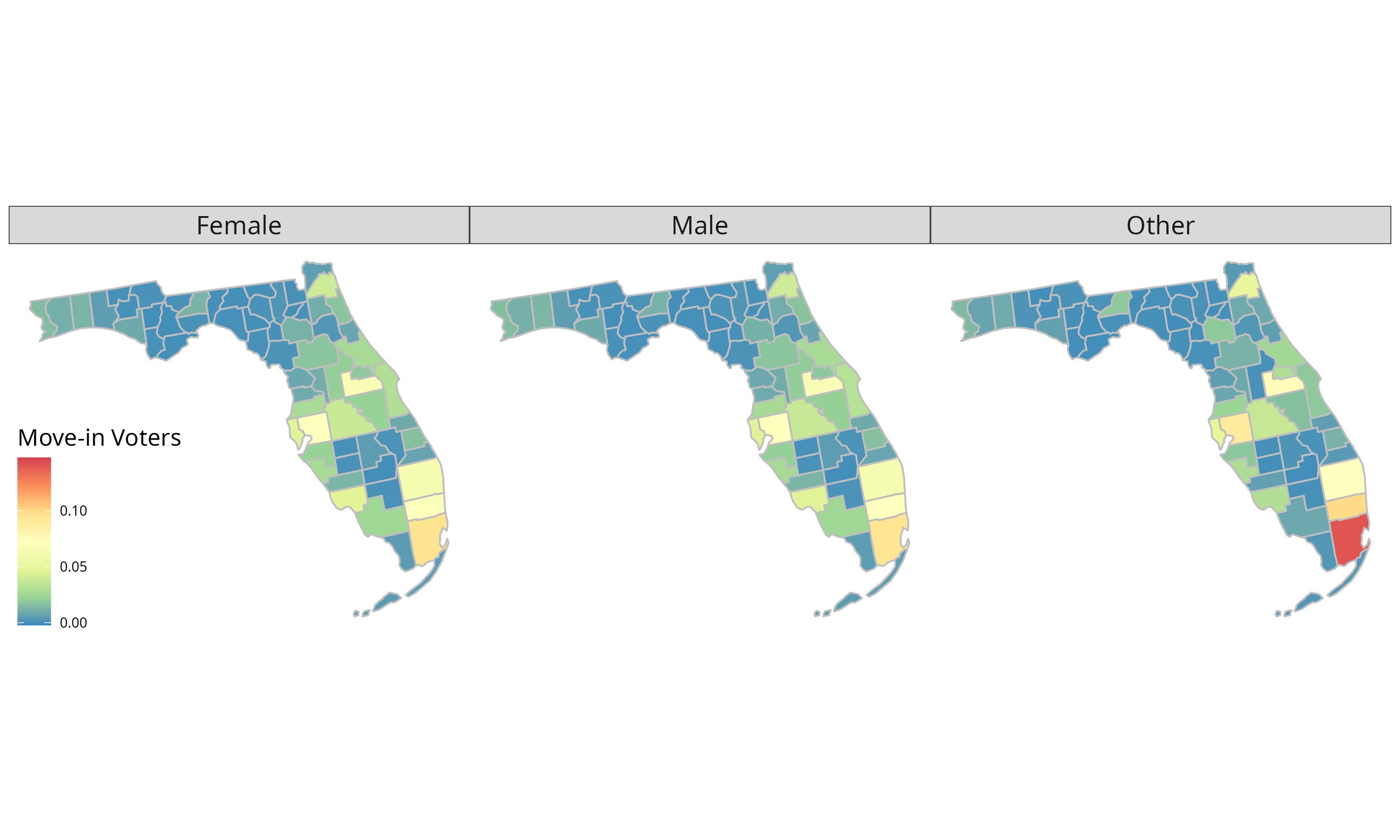}
  \caption{Ratio of move-in voters by gender.\label{fig:map-gender-in}}
\end{figure}

\begin{figure}[H]
  \centering
  \includegraphics[width=\linewidth]{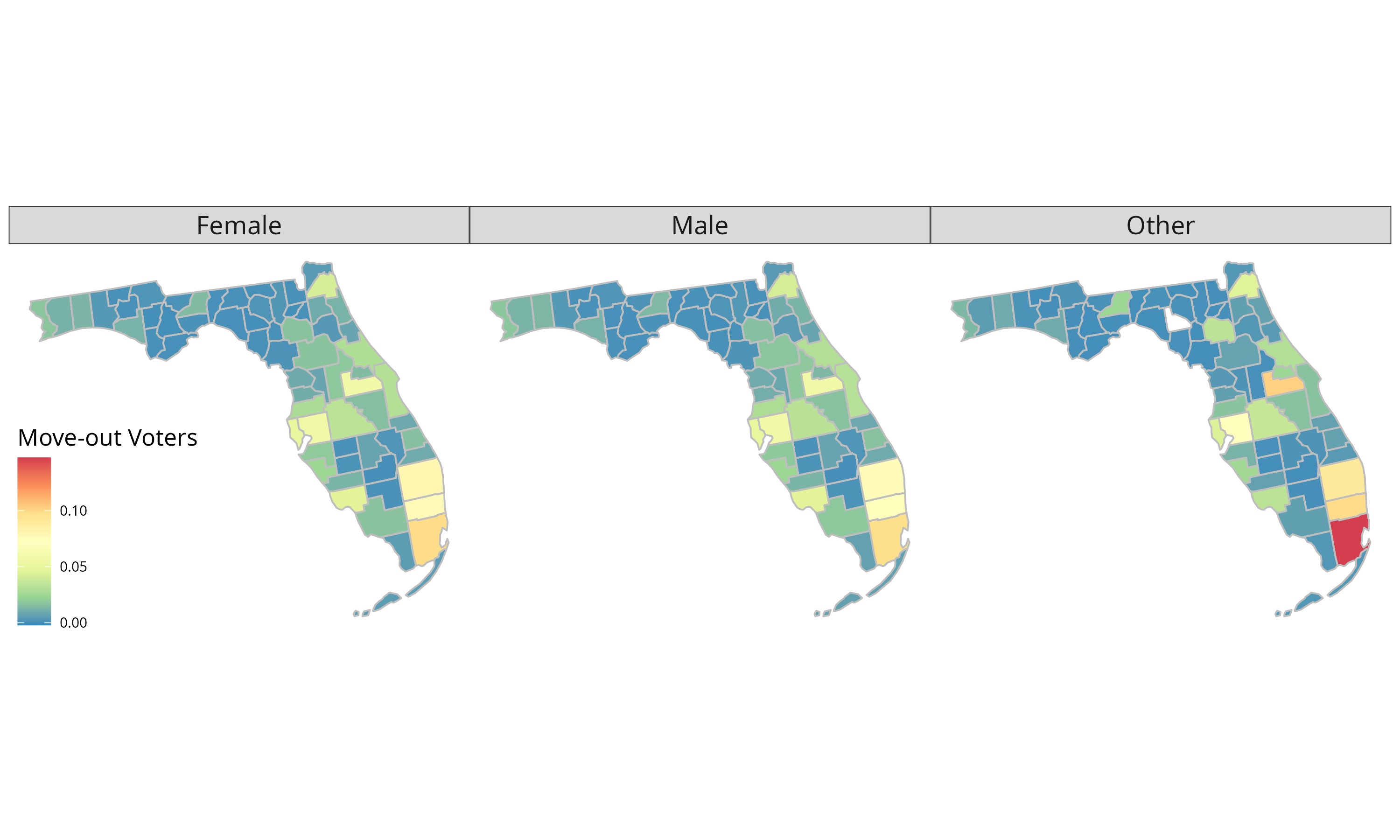}
  \caption{Ratio of move-out voters by gender.\label{fig:map-gender-out}}
\end{figure}

If we stratify the voter samples by gender,
we can see a similar pattern as in \cref{subsec:geo-race},
i.e.\ most of the migration happens in the urban areas.
From \cref{tab:summary_inoutnet} we know that
the move-in and move-out voters are relatively balanced in gender,
but this pattern persists even at the county level,
as can be seen in \cref{fig:map-gender-in,fig:map-gender-out}.

\begin{figure}[H]
  \centering
  \includegraphics[width=\linewidth]{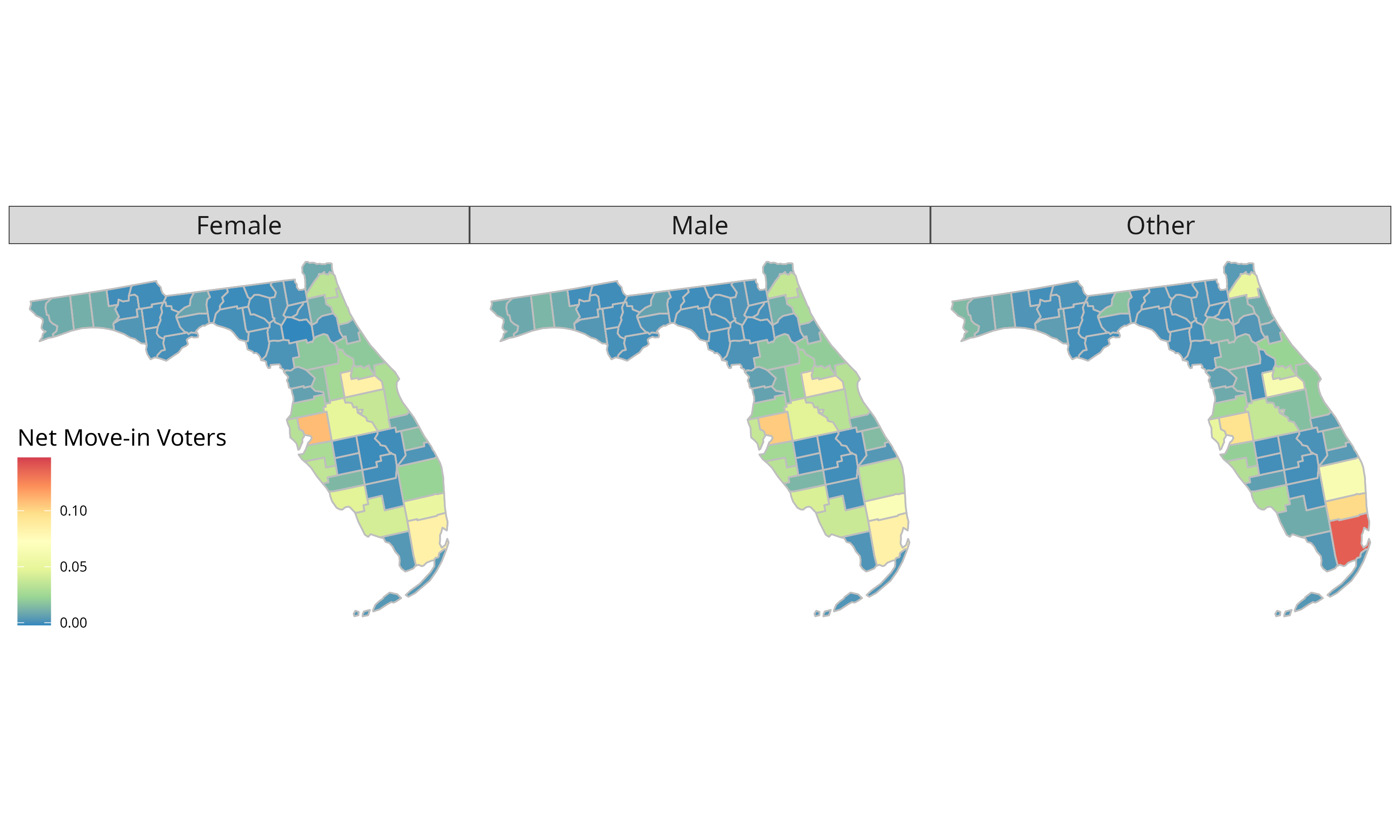}
  \caption{Ratio of net move-in voters by gender.\label{fig:map-gender-net}}
\end{figure}

What's more interesting is the pattern of the unknown gender \cref{fig:map-gender-net},
as it displays a similar geographic distribution
as in the racial minorities in \cref{subsec:geo-race}.
The voters with unknown gender tend to move into the metropolitan areas
with more concentration, and less spread out as male and female voters.

As seen in \cref{tab:summary_inoutnet} discussed in \cref{sec:methods},
there appears to be a non-trivial proportion (11.9\%) of voters
with unknown gender among the net move-in voters between 2017 and 2022.
The Florida Voter registration files,
provided by Florida Division of Elections,
did not explain why some voters have unknown gender.
According to the voter registration form in Florida\footnote{
  \url{https://files.floridados.gov/media/704789/1s-2040-form-ds-de-39-eng-fillable-20251017.pdf}
},
there is a checkbox for the registrant to indicate their gender (M/F),
and we can only assume that those voters without specified gender
did not fill this checkbox.
This is the only reasonable explanation that could result in missing gender
information for the voters.

The missing gender information could plausibly result from either
(1) registrants inadvertently skipping the gender checkbox on the form, or
(2) registrants intentionally choosing not to select
either of the two binary options provided.
We do not have sufficient information to distinguish between
these two possibilities.
However, we note that the completion rates for other fields
(age, race, party, etc.)\ are generally high,
which may suggest that at least some of the missing responses
are intentional rather than accidental.

One possible---though speculative---explanation is that
some of these voters chose not to identify with a binary gender category.
A Gallup survey \citep{gallup2025} reports that
the proportion of U.S.\ adults identifying as LGBTQ+
has been increasing over time, reaching 9.3\% in 2024,
with notably higher rates among younger cohorts
(e.g., 22.7\% among Gen~Z adults in 2022).
Given that nearly half of the net move-in voters
fall within the 18--34 age group (\cref{tab:summary_inoutnet}),
changing attitudes toward gender identity among younger generations
could be one contributing factor.
It must be emphasized, however, that this remains a hypothesis;
the available data do not allow us to draw definitive conclusions,
and further investigation with more detailed survey data
would be needed to test this hypothesis.

\subsection{Geographic Distribution of Voter Migration by Age Group}\label{subsec:geo-age}

\begin{figure}[H]
  \centering
  \includegraphics[width=\linewidth]{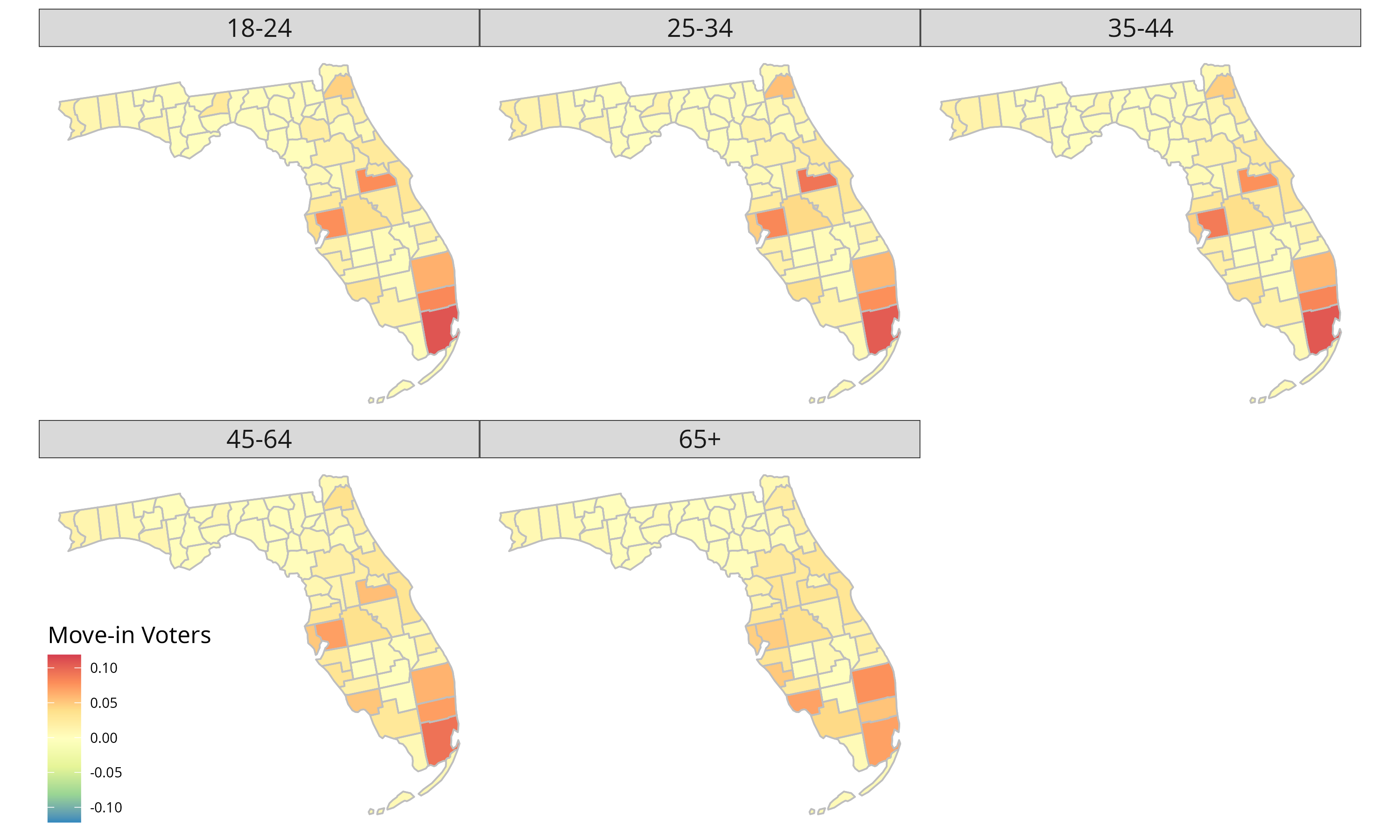}
  \caption{Ratio of move-in voters by age group.\label{fig:map-age-in}}
\end{figure}

\begin{figure}[H]
  \centering
  \includegraphics[width=\linewidth]{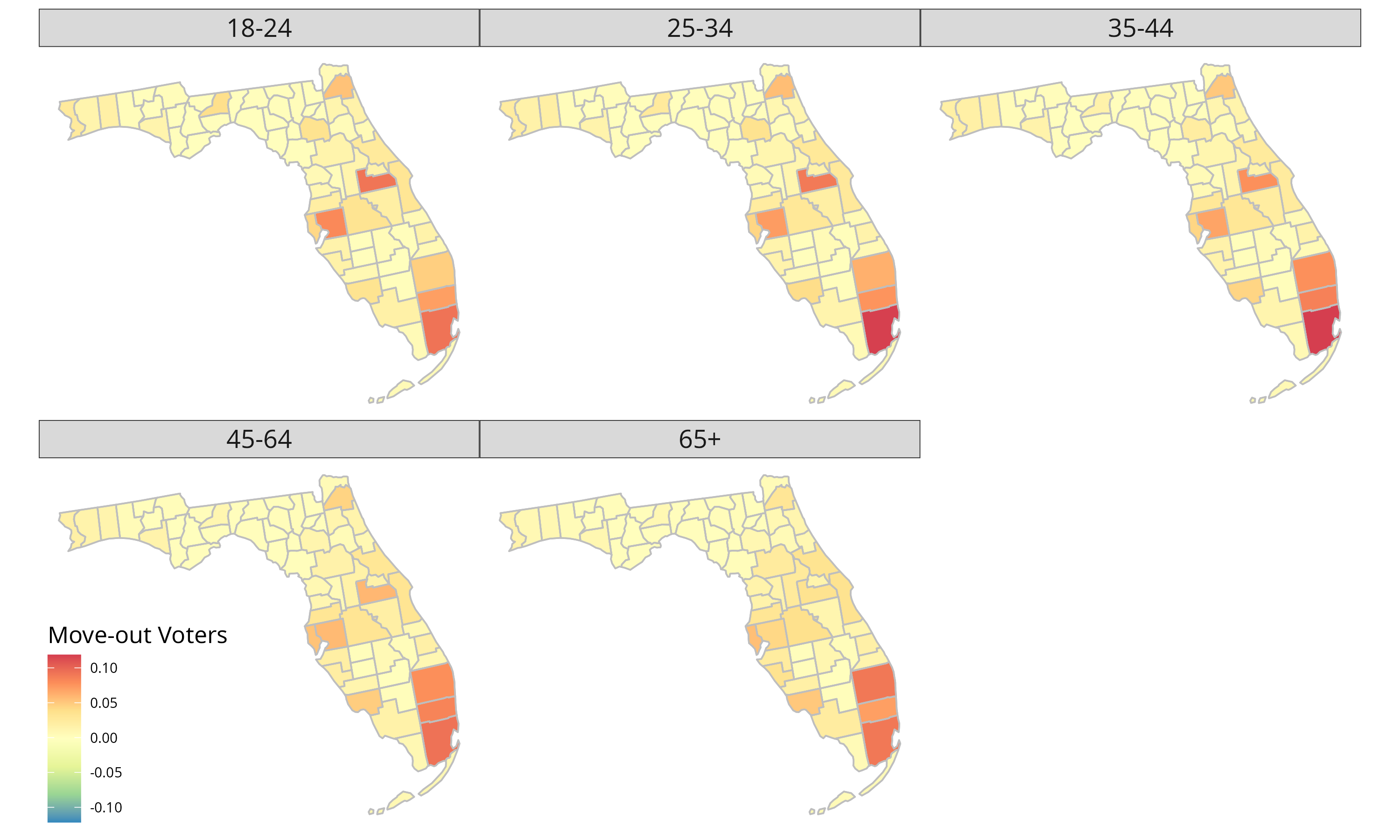}
  \caption{Ratio of move-out voters by age group.\label{fig:map-age-out}}
\end{figure}

We have already seen in \cref{sec:methods} that there is a large
outflow of voters in the 65+ age group during this period,
which is quite counter-intuitive given Florida's reputation
as a popular retirement destination.
Now, we can further visualize the geographic distribution
of the migration patterns by age group.
From \cref{fig:map-age-in,fig:map-age-out},
we can see that the migration patterns are quite similar
to what we have observed in \cref{subsec:geo-race,subsec:geo-gender},
i.e.\ most migration occurred in the urban areas.
However, there are significantly fewer move-in voters
in the 65+ age group compared to other age groups,
especially in the Miami-Dade, Hillsborough, and Orange counties.
In fact, the county with the largest inflow of the 65+ group
is the Palm Beach County.
Moreover, the 65+ group is also more spread out than other age groups,
for example, they moved into the neighboring counties of
Orange and Hillsborough more so than other age groups.

If we also look at the move-out voters in \cref{fig:map-age-out},
we can see that most of the people who moved out from 2017 to 2022
are mostly concentrated in the several urban counties,
but again with the 65+ group moving out from a wider range of counties.

\begin{figure}[H]
  \centering
  \includegraphics[width=\linewidth]{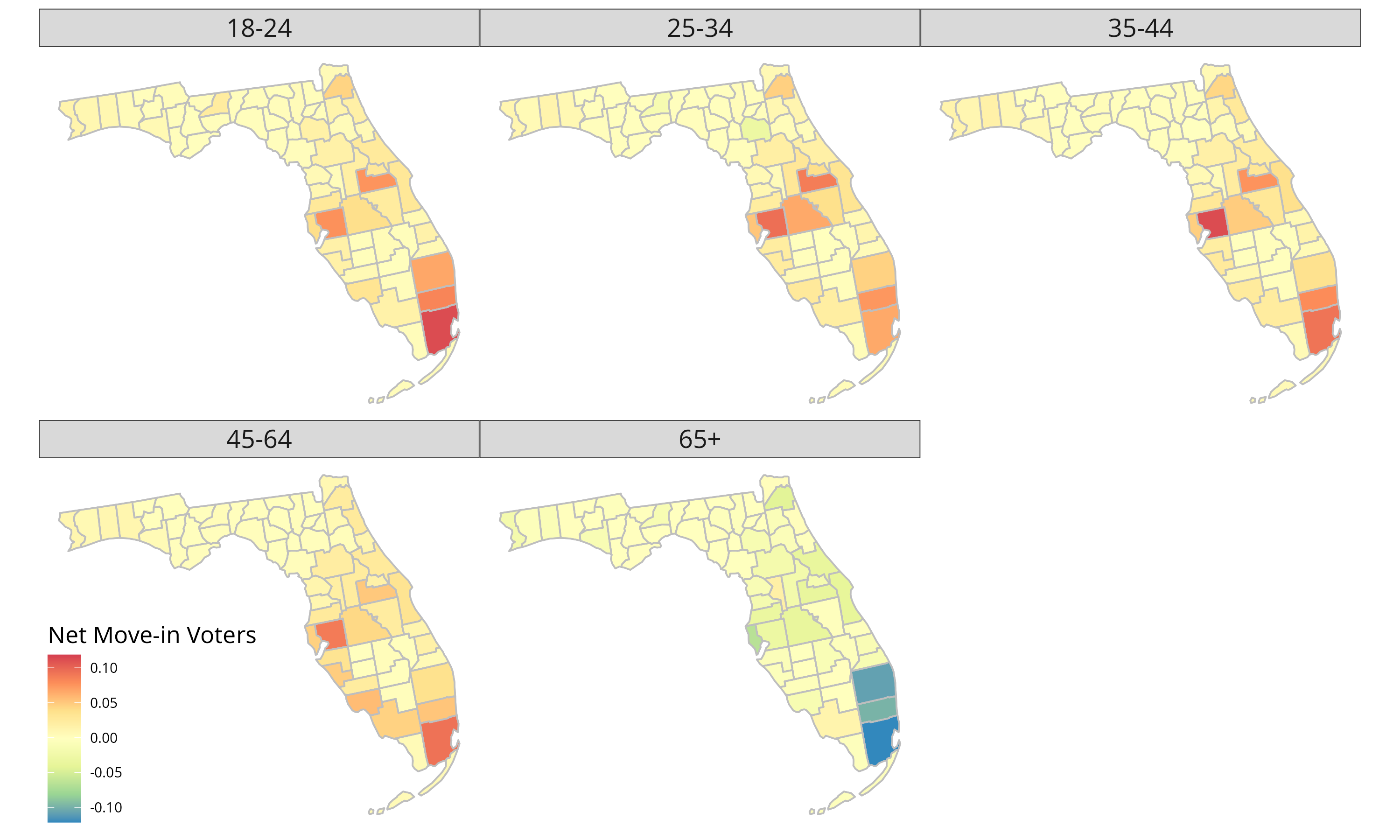}
  \caption{Ratio of net move-in voters by age group.\label{fig:map-age-net}}
\end{figure}

Finally, if we look at the net move-in voters in \cref{fig:map-age-net},
the net inflow pattern confirms our previous observation in
\cref{tab:summary_inoutnet} that most of the counties saw
a net decline in the registered voters in the 65+ group,
while all other age groups were having a net increase.

The retirement destination counties map provided by USDA \citep{usda2025}
clearly shows that 63\% of the retirement destination
counties are in rural areas,
and more specifically in Florida, these counties are mostly located
in the entire Florida other than the urban counties
(e.g. Miami-Dade, Broward, Palm Beach, Hillsborough, Orange, Duval, etc.).
The USDA study indicates that the retirement-destination counties
``are those in which the number of residents ages 55-74 increased by
at least 15 percent from 2010 to 2020 because of migration.''
Notice that the USDA definition of migration covers a longer period of time
(2010-2020) than this study (2017-2022),
while also covers a bigger group of population (55-74 versus 65+).

The trend for retirees moving into rural counties is also well studied
in the literature (for example \citep{serow2001}).
Combined with the USDA study on retirement destinations in the prior decade,
the fact that a large number of voters above 65 moved out of not just
urban counties but also rural counties
makes this trend even more surprising.
This is perhaps due to the large influx of new residents into Florida
since the outbreak of the Covid pandemic,
which has driven up the cost of living in Florida,
making it less affordable for retirees to stay.

Given the limitation of the voter registration data,
we can only identify the migration patterns of voters between 2017 and 2022,
and we cannot identify the exact reasons for the new migration patterns
for the retirees.
There are two potential issues specifically worth discussing
regarding the migration patterns of the 65+ group\footnote{
  Those are suggested by two anonymous reviewers,
  and I thank them for the insightful comments and suggestions.
}:
(1) how would the mortality of the 65+ group affect the migration patterns?
(2) how would the intra-state relocations between counties affect the migration patterns?

\begin{figure}[H]
  \centering
  \includegraphics[width=\linewidth]{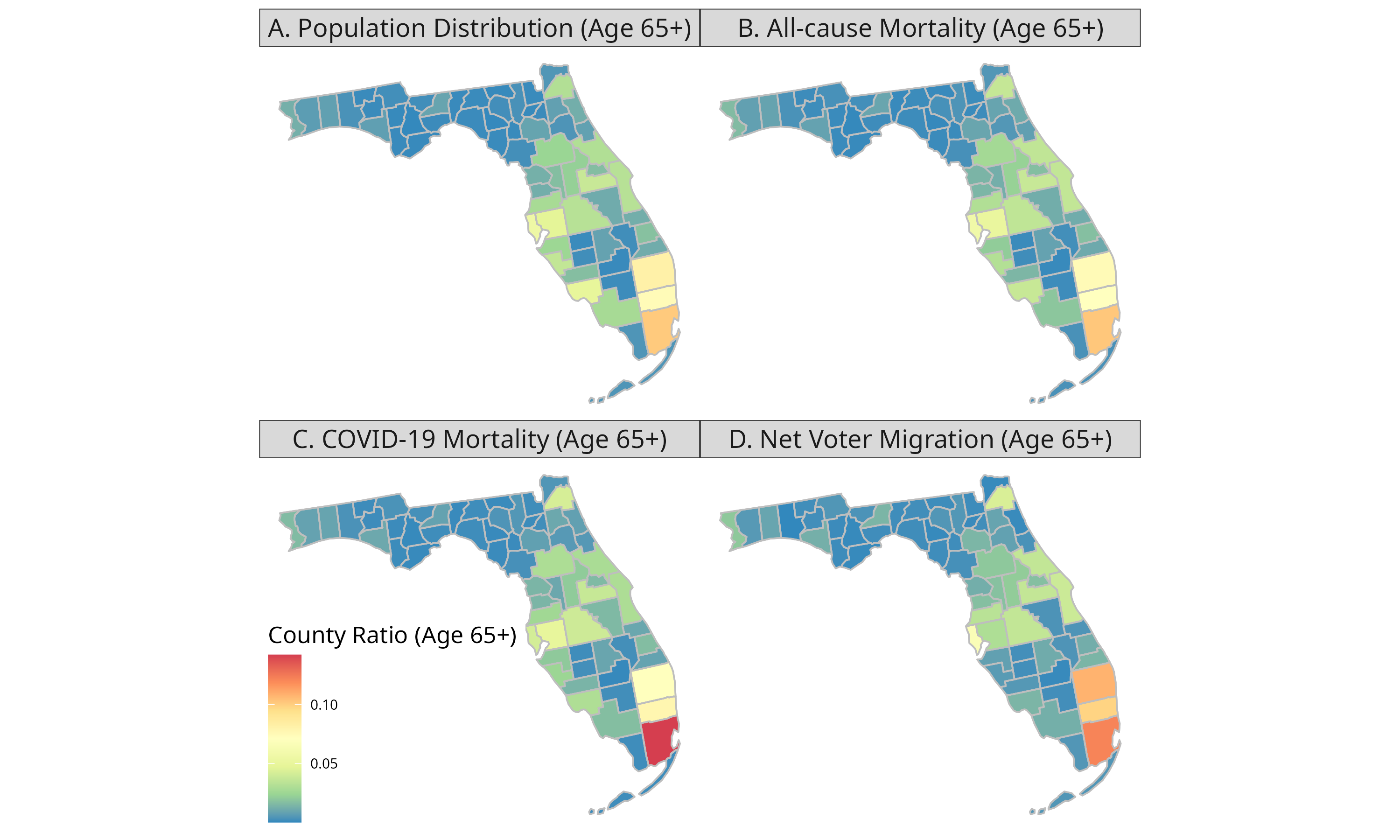}
  \caption{Ratio of populations, mortality, and net voter migration for Age 65+.\label{fig:map-pop-mort}}
\end{figure}

For the first issue, I obtained the population data from the American Community Survey (ACS)
5-year estimates between 2018 and 2022,
the mortality data from the CDC WONDER database for the same period (by all causes and Covid),
and calculated the ratio of populations, mortality, and net voter migration for the 65+ group
in \cref{fig:map-pop-mort}.
From the map, we can see that the all-cause mortality (Panel B) is in proportion to the
population size across counties (Panel A),
whereas the Covid mortality (Panel C) shows a disproportionately higher mortality rate
in the Miami-Dade county.
By comparing the net voter migration (Panel D) with the population and mortality patterns,
we can see that there's clearly a net out-migration of voters in the 65+ group in Miami-Dade county,
Broward county, and Palm Beach county, which cannot be explained by the population size or mortality patterns.
Note that all but a few counties in Florida have a net out-migration of voters in the 65+ group,
and thus \cref{fig:map-pop-mort} plots the absolute values of the net voter migration
for easier comparison with the population and mortality patterns,
as opposed to the raw negative values shown in \cref{fig:map-age-net}.

\begin{figure}[H]
  \centering
  \includegraphics[width=\linewidth]{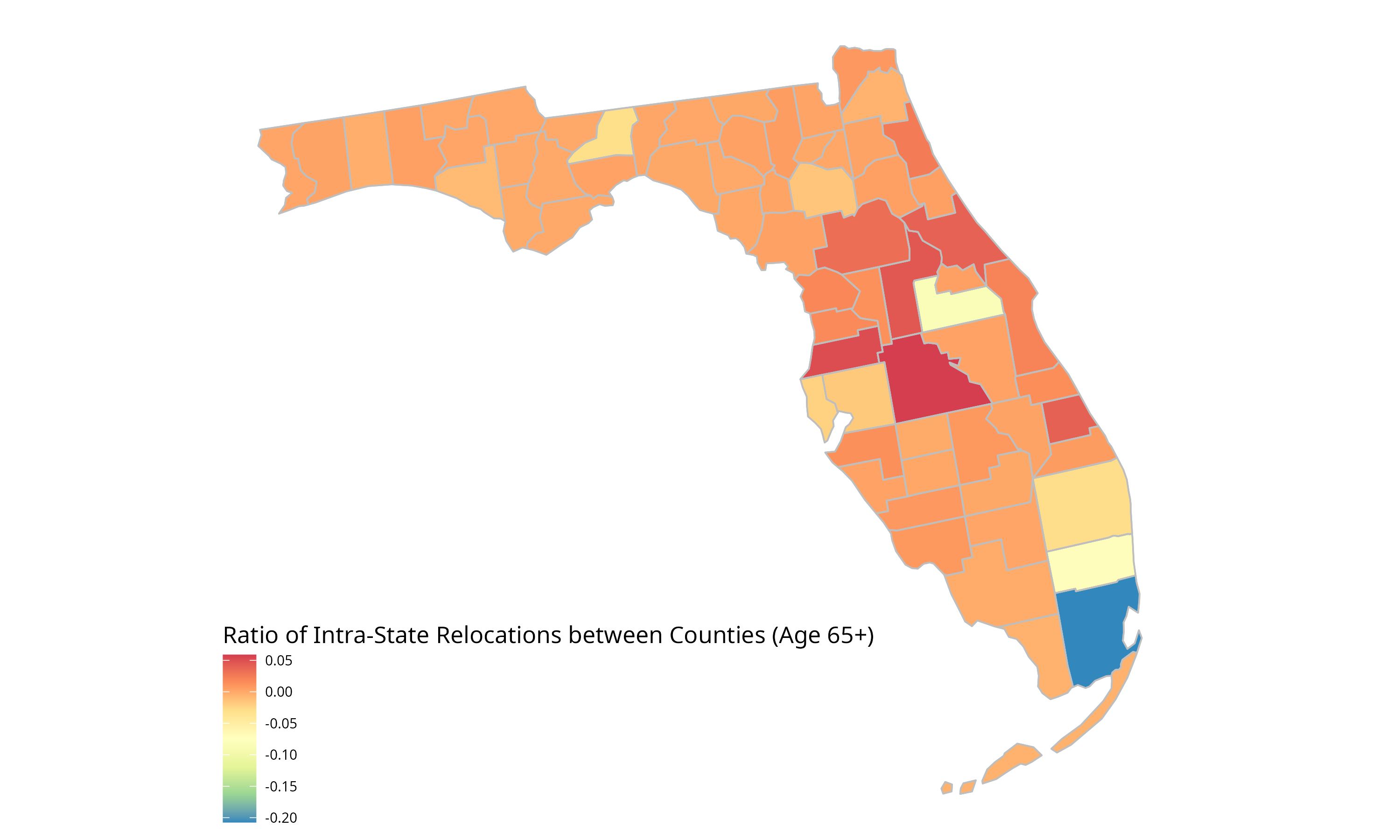}
  \caption{Ratio of intra-state relocations between counties for Age 65+.\label{fig:map-intermove}}
\end{figure}

To further illustrate the migration patterns of the 65+ group,
I also calculated the ratio of intra-state relocations between counties for the 65+ group
in \cref{fig:map-intermove}.
The intra-state relocations are identified by the voters who appear in both 2017 and 2022 files,
but with different county FIPS codes in the two files.
From the map, we can see that the intra-state relocations for the 65+ group
are mostly concentrated in the urban areas, such as Miami-Dade, Broward, Palm Beach,
Hillsborough, and Orange counties.
This is consistent with the patterns we have observed in \cref{fig:map-age-net}
for the net move-in voters in the 65+ group,
but the intra-state relocations further confirm the scale of
65+ voters moving out of the metropolitan areas.
On the other hand, the central counties neighboring the Orange and Hillsborough counties
are clearly favored by the 65+ voters who relocate within the state.
Such migration patterns demonstrate that the retirees are not just moving out of the urban areas,
but also moving into the neighboring counties of the urban areas,
which are still relatively more affordable than the urban areas,
while still being close enough to the urban areas for better access to
healthcare and other services.
It is worth noting that this pattern of short-distance relocation
to adjacent counties appears more consistent with
a redistribution of residents within existing metropolitan regions,
where infrastructure and services are already in place,
than with outward urban sprawl into undeveloped areas.

\subsection{Geographic Distribution of Voter Migration by Party Affiliation}

Last but not least, we can also document the migration patterns
by party affiliation across Florida counties.
This aspect is perhaps even more interesting given the changing political
landscape in Florida in recent years,
especially with the Republican Party solidifying its dominance
in the state.
The fact that our dataset comes from voter registration files
makes it particularly suitable for this analysis.

Florida was almost exclusively Democratic,
before turning into Republican in 1952.
The influx of immigrants with diverse backgrounds
made it a swing state, especially in the 2000 presidential election,
but it has become more reliably Republican in recent years \citep{270towin2024}.
One of the main reasons that led to Florida losing its status
as a battleground state is the large number of Republicans
moving into Florida than Democrats \citep{anderson2024}.

\begin{figure}[H]
  \centering
  \includegraphics[width=\linewidth]{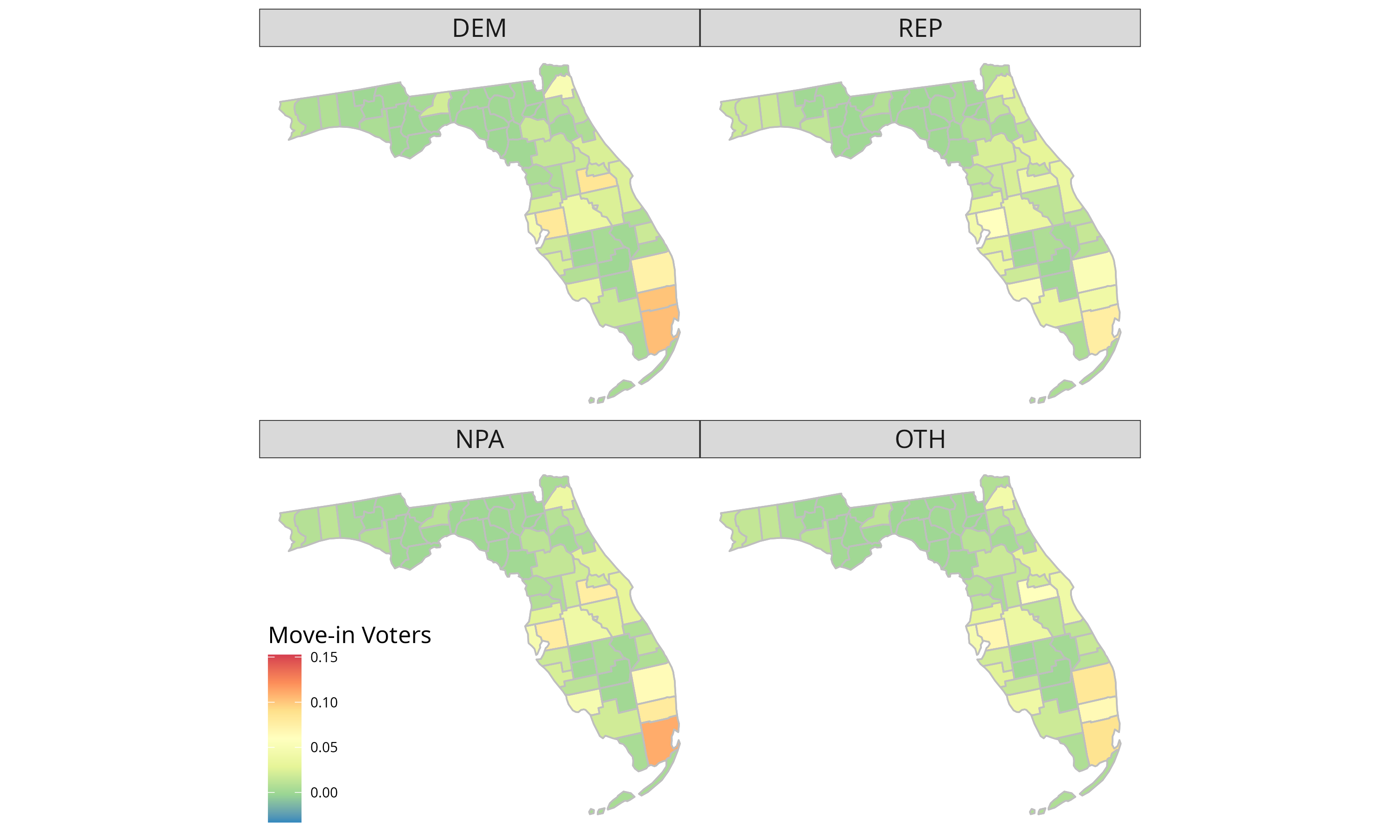}
  \caption{Ratio of move-in voters by party affiliation.\label{fig:map-party-in}}
\end{figure}

We can clearly see this trend in \cref{fig:map-party-in},
where the majority of the move-in voters who identify themselves as
Democrats are in Miami-Dade, Broward, Hillsborough, and Orange counties.
Republicans, on the other hand,
are more spread out across all counties,
with less concentration in the urban areas.

\begin{figure}[H]
  \centering
  \includegraphics[width=\linewidth]{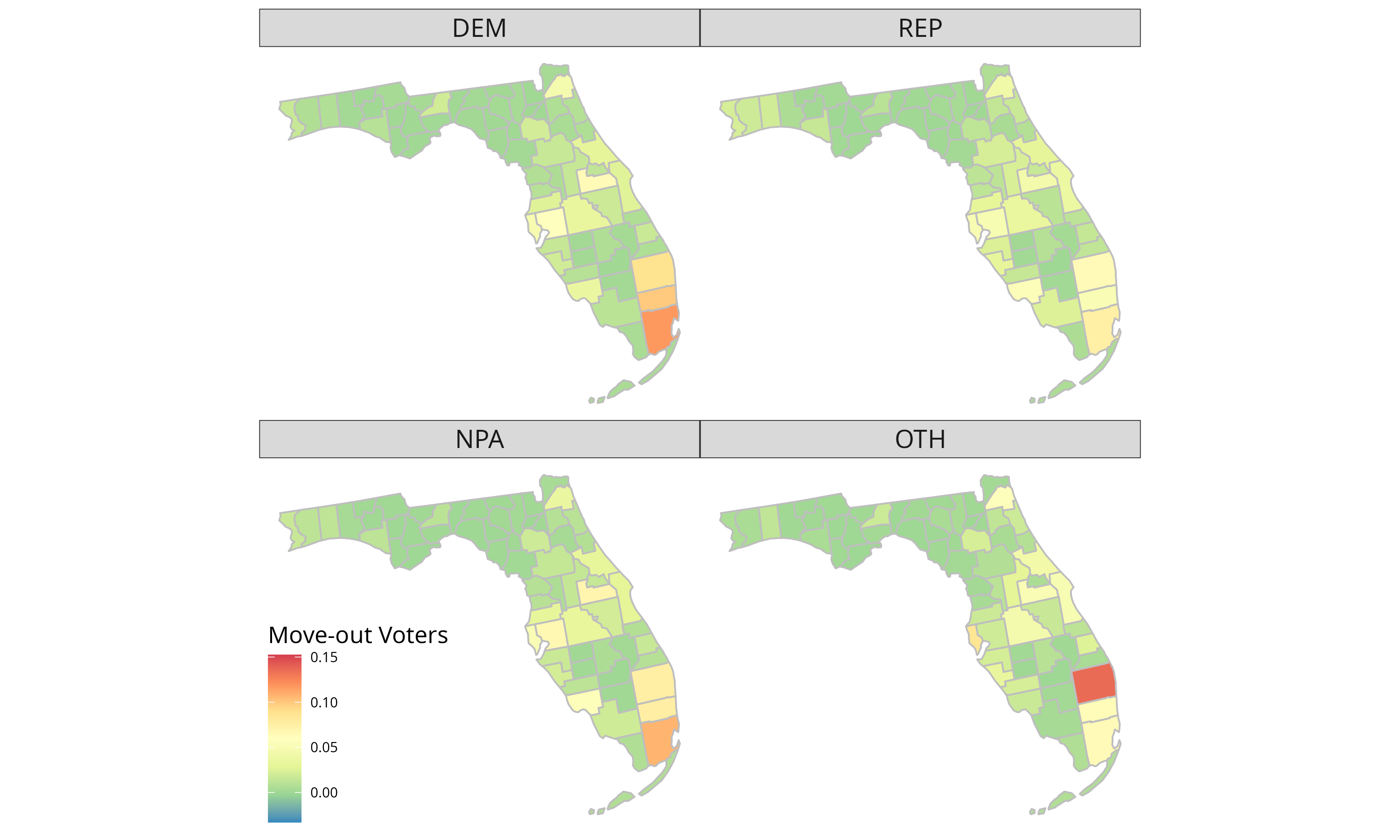}
  \caption{Ratio of move-out voters by party affiliation.\label{fig:map-party-out}}
\end{figure}

If we look at the move-out patterns in \cref{fig:map-party-out},
the Democrats are mostly leaving the Miami area
(Miami-Dade, Broward, Palm Beach).
The non-affiliated voters (NPA) are also leaving Miami region
with a similar pattern to Democrats,
but the voters who register as members of other parties
were primarily leaving Palm Beach County.
The Republicans, however, seem to move out from a wider range of counties,
with less concentration.

\begin{figure}[H]
  \centering
  \includegraphics[width=\linewidth]{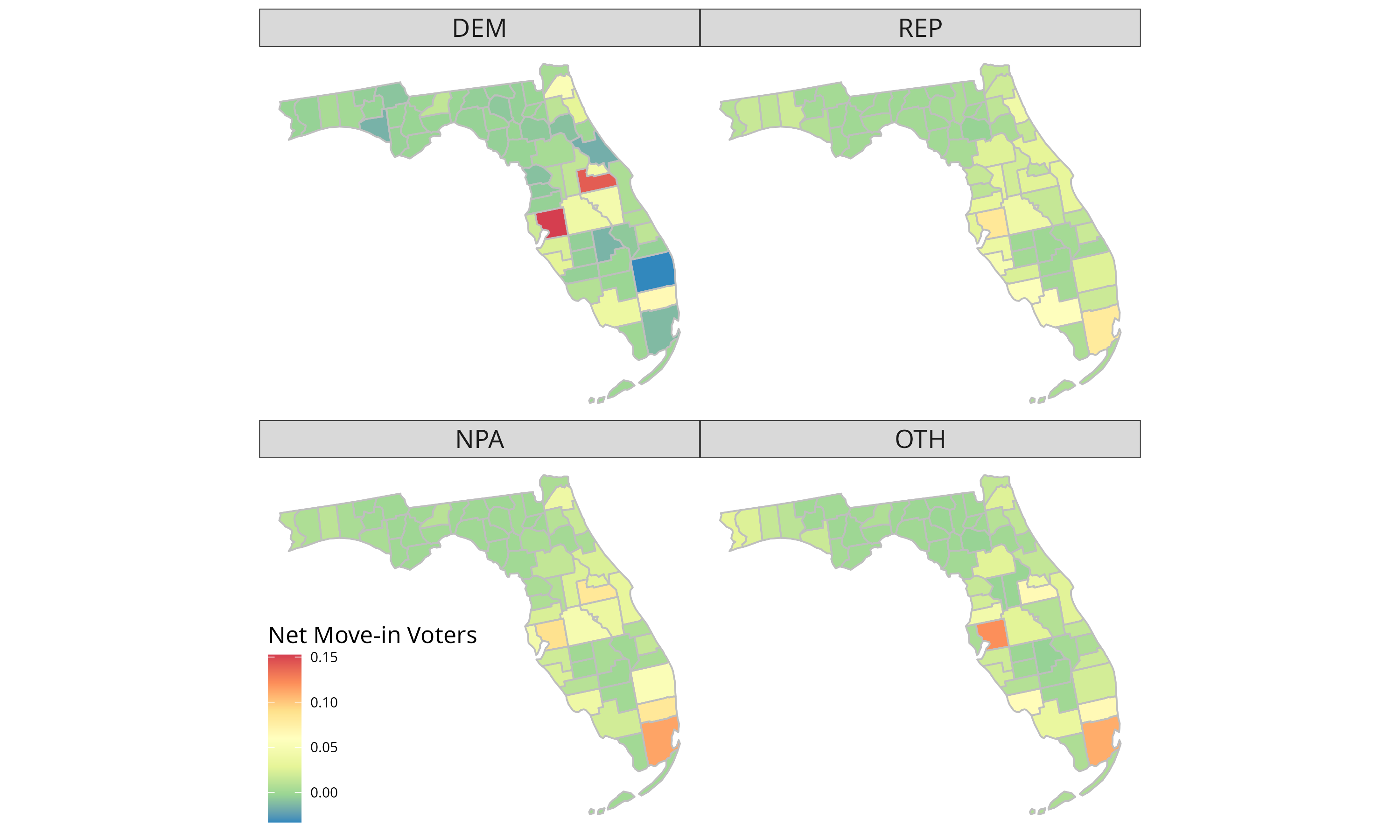}
  \caption{Ratio of net move-in voters by party affiliation.\label{fig:map-party-net}}
\end{figure}

Finally, if we look at net move-in voters in \cref{fig:map-party-net},
we can see that most counties saw a net increase of
Republican voters, as well as NPAs and OTHs.
Democrats, on the other hand, are mostly moving out of Florida,
especially in Palm Beach and Miami-Dade counties,
as well as in many rural counties.
They do, however, have a very concentrated net increase in Hillsborough
and Orange Counties.

\section{Conclusions}\label{sec:conclusion}

In this paper, I analyze the migration patterns of voters in Florida
between 2017 and 2022 using Florida voter registration data.

The analysis reveals that although the non-Hispanic White population is still
the majority, more minorities moved into Florida
than moving out between 2017 and 2022.
Younger voters are also more likely to move into Florida,
especially those aged between 18-24,
while 25-34 group shows the smallest inflow.
Surprisingly, retirees (aged 65+) show a significant net outflow,
which contradicts the common perception of Florida as a popular
retirement destination.
There is also a non-trivial number of voters with unknown gender,
who did not select the male/female options on the registration form.
While we cannot determine the exact reasons, one possible hypothesis is that
some of these voters are gender non-conforming.
This group had a notable net inflow into Florida between 2017 and 2022.
Finally, Republicans are more likely to move into Florida
than Democrats during this period,
with over half of the voters not affiliated with either party.

The further geographic analysis of migration patterns
shows that net move-in voters are concentrated
in the metropolitan areas of Florida,
such as Miami, Tampa, Orlando, and Jacksonville,
especially for minorities (Asian and Hispanic) and younger voters.
We also observe a significant net outflow
of non-Hispanic White voters in Broward and Palm Beach Counties,
and of Black voters in Miami-Dade County.
There is a non-trivial number of voters with unknown gender
moving into Miami, Tampa, Orlando, and Jacksonville,
who may---though this remains speculative---include individuals
who chose not to identify with a binary gender category.
In terms of age, the working-age population generally moves into
urban counties, while retirees (aged 65+) tend to move out
of most counties in Florida, particularly in the Miami area.
Finally, Republicans are more likely to move into
most of Florida,
while Democrats tend to move out of most areas
other than Tampa and Orlando.

The voter migration analysis in this paper sheds light
on the demographic and geographic patterns
of population movement in Florida between 2017 and 2022.
These findings have important implications for understanding
the political landscape of Florida,
as well as for policymakers and political strategists
seeking to engage with the state's diverse electorate.

\newpage
\addcontentsline{toc}{section}{References}
\printbibliography

\newpage

\end{document}